\documentclass[11pt,a4paper]{article}
\usepackage{jheppub}
\usepackage[latin9]{inputenc}
\usepackage{amsmath}
\usepackage{amssymb}
\usepackage{setspace}
\usepackage{graphicx}
\usepackage{babel}

\title{Baryonic Vortex Phase and Magnetic Field Generation in QCD with Isospin and Baryon Chemical Potentials}

\author[a,b]{Zebin Qiu\,
}
\emailAdd{qiuzebin@keio.ac.jp}

\author[a,b,c]{and Muneto Nitta\,
}
\emailAdd{nitta@phys-h.keio.ac.jp}
\affiliation[a]{%
Research and Education Center for Natural Sciences, Keio University, 4-1-1 Hiyoshi, Yokohama, Kanagawa 223-8521, Japan}
\affiliation[b]{
Department of Physics, Keio University, 4-1-1 Hiyoshi, Yokohama, Kanagawa 223-8521, Japan}
\affiliation[c]{
International Institute for Sustainability with Knotted Chiral Meta Matter(SKCM$^2$), Hiroshima University, 1-3-2 Kagamiyama, Higashi-Hiroshima, Hiroshima 739-8511, Japan
}
\abstract{
We propose a novel baryonic vortex phase in low energy dense QCD with finite baryon and isospin chemical potentials. 
It is known that the homogeneous charged pion condensate emerges as a ground state at finite isospin chemical potential, and therein arises the Abrikosov vortex lattice with an applied magnetic field.
We first demonstrate that a vortex with the same quantized magnetic flux 
as the conventional Abrikosov vortex,
carries a baryon number captured by
the third homotopy group of Skyrmions, 
once we take into account a modulation of the neutral pion inside the vortex core.
Such a vortex-Skyrmion state is therefore dubbed the baryonic vortex.
We further reveal that when the baryon chemical potential is above a critical value, 
the baryonic vortex has negative tension measured from the charged pion condensation.
It implies that the phase, in which such vortices emerge spontaneously without an external magnetic field, would take over the ground state at high baryon density. 
Such a new phase contributes to the comprehension of QCD phase diagram and relates to the generation of magnetic fields inside neutron stars. 
}

\begin{document}

\maketitle

\section{Introduction}
% low energy dense QCD
Quantum Chromodynamics (QCD) phase at finite density is a long-standing hot issue in nuclear physics~\cite{Eichmann:2016yit, Schmidt:2017bjt, Fischer:2018sdj, Rothkopf:2019ipj, Guenther:2020jwe, Braguta:2023yhd, Adam:2023cee}.  
In the interested case of 2-flavor ($N_f=2$) QCD with $u$ and $d$ quarks, the finite density effects can be captured via the isospin chemical potential $\mu_I=(\mu_u - \mu_d)/2$ and the baryon chemical potential $\mu_B=(\mu_u + \mu_d)/2$.
Finite $\mu_{I,B}$ coexist in the early universe with lepton flavor asymmetry, astrophysical objects such as neutron stars, and needless to say, 
heavy-ion collisions.
Despite immense research interest over decades, the phase structure of dense QCD is still challenging to fully understand given the formidable theoretical study on strongly correlated systems from first principles as well as the technical difficulty typified by the sign problem of Lattice QCD (LQCD) at finite baryon density.
However, in the low energy regime, one can resort to effective field theories (EFTs)~\cite{Holt:2014hma,Hammer:2019poc,Drischler:2021kxf}. 
Even within long wavelength physics, novel hadronic phases with distinct topology have been discovered, which are the main interest of the present paper.
The infrared dynamics of QCD is based on the confinement and the chiral symmetry breaking, which are common grounds shared by both 2-flavor and 3-flavor QCD. 
A specific formalism is the chiral perturbation theory (ChPT)~\cite{Gasser:1983yg,Scherer:2002tk} in terms of pion fields which arise as Nambu-Goldstone (NG) bosons as a result of chiral symmetry breaking.
The topology in ChPT framework is encoded in the conserved Goldstone-Wilczek (GW) current $j_B$~\cite{Goldstone:1981kk}, which could be expressed as a 3-form constructed with the $SU(N_f)$ current. 
The coupling of such current with gauge fields via the triangle anomaly is incarnated in the Wess-Zumino-Witten (WZW) term~\cite{Witten:1983tw}, where the gauge field could be either the electromagnetic $A_\mu$ or the baryonic $A_\mu^B$, 
and 
$\mu_I$ and $\mu_B$ are nothing but zeroth components of $A_\mu$ and $A_\mu^B$, respectively. 

% baryon chemical potential
QCD phase structure at finite $\mu_B$ becomes especially intriguing when a magnetic field is present, as reviewed in Refs.~\cite{Kharzeev:2015kna,Miransky:2015ava,Andersen:2014xxa,Yamamoto:2021oys,Cao:2021rwx,Iwasaki:2021nrz}. 
Especially, the magnetic catalysis~\cite{Klimenko:1990rh} is considered a general effect at zero temperature, {\it i.e.}, magnetic field either induces the quark condensate or enhances an already existent condensate. 
More recently, an opposite tendency at finite temperature is uncovered as the inverse magnetic catalysis~\cite{Bruckmann:2013oba}, which is more pronounced when the temperature is approaching the critical temperature of the crossover~\cite{Aoki:2006we} between hadronic and quark phases.
In reality, strong magnetic fields are prevalent in aforementioned systems like magnetars~\cite{Duncan:1992hi,Turolla:2015mwa,Kaspi:2017fwg} and quark gluon plasma~\cite{Shuryak:1978ij,Skokov:2009qp,McLerran:2013hla}.
In LQCD, a magnetic field makes the sign problem moderated~\cite{Fukuda:2013ada} or even bypassed~\cite{Brauner:2019rjg}. 
All these facts keep steadfast attention to studying QCD with the magnetic field as a probe. 
For the record, we nominate magnetically induced states in quark matter, chiral magnetic spiral~\cite{Basar:2010zd}, and dual density wave~\cite{Nakano:2004cd}, for interested readers.

More pertinent to the hadronic sector that we focus on are the proposals of a neutral pion $\pi^0$ domain wall~\cite{Son:2007ny} and its stacking offshoot, chiral soliton lattice (CSL)~\cite{Eto:2012qd,Brauner:2016pko}.
In short, CSL is a result of the dimensional reduction that occurs when an axial magnetic field breaks the symmetry $SU(2)$ further into $U(1)$. 
Meanwhile the still conserved topological charge decreases the free energy through the WZW term and breaches the vacuum when the magnetic field is above critical value $B\geq B_{\text{CSL}}=16\pi m_\pi f^2_\pi / e \mu_B$ (with $m_\pi$ the effective pion mass and $f_\pi$ the pion decay constant).
In the wake of such studies, there has been a trend of thought that the ground state of dense matter under strong $B$ could be pure pionic~\cite{Brauner:2018mpn}, in contrast to familiar baryonic states. 
A similar CSL composed of the $\eta'$ meson 
(or $\eta$ meson for the two-flavor case) 
\cite{Huang:2017pqe,Nishimura:2020odq,Chen:2021aiq} 
or its decomposition to a non-Abelian CSL \cite{Eto:2021gyy,Eto:2023rzd} 
was found to be the ground state under rapid rotation.

% further studies 
In view of the underlying topology, from the commonly defined GW current $j_B$, one can manifest that the two kinds of states feature distinguished homotopy, $\pi_1[U(1)] \simeq {\mathbb Z}$ for the $\pi^0$ domain wall and $\pi_3[SU(2)] \simeq {\mathbb Z}$ for the baryon arising as a Skrymion~\cite{Adkins:1983ya}. 
In the latter case, the physical meaning of $j_B$ is nothing but the baryon charge current.
The conserved baryon number is sustained in a finite size system guaranteed by the quartic Skyrme term that overcomes Derrick's scaling law \cite{Derrick:1964ww}. 
Along this line, CSL could be further put into the same framework with crystalline baryons under $B$ via the simplest extension of ChPT, {\it{i.e.}}, the Skyrme model~\cite{Skyrme:1961vq,Zahed:1986qz}.   
Interestingly, a phase diagram of CSL and baryonic crystal is established with parameters of baryon area density and magnetic field~\cite{Chen:2021vou}, stipulating the CSL lives in the region with lower density and higher magnetic field.
Furthermore, when both magnetic field and density are high, a more sophisticated ground state has been found as the domain-wall Skyrmion~\cite{Eto:2023lyo,Eto:2023wul} (see also Ref.~\cite{Eto:2023tuu} for a rotation counterpart), which is essentially a topological lump 
supported by $\pi_2(S^2) \simeq {\mathbb Z}$
%(the jargon whose counterpart in condensed matter physics is the ``magnetic Skyrmion'') 
lodged on top of a domain wall 
\cite{Nitta:2012wi,Nitta:2012rq,Gudnason:2014nba,Gudnason:2014hsa}, 
understood in view of the decomposition $SU(2)/U(1)\simeq S^2$. 
Moreover, when the density becomes sufficiently high for $\eta$ mesons to come into play, we have proposed in preceding work~\cite{Qiu:2023guy} the idea of mixed soliton lattice, which comprise $\pi^0$ and $\eta$ mesons while carrying lower energy than their separate lattices. Our results lead to the conjecture, for the first time, a quasicrystal in QCD context. 
These studies substantiate the rich phase structure of hadronic matter with $\mu_B$ and $B$.

% $\mu_I$
Yet possibilities of hadronic phases could be even more opulent when the isospin chemical potential $\mu_I$ is taken into account.
It has been known for a long time that a Bose-Einstein condensation of charged pions $\pi^\pm$~\cite{Ruck:1976zt, Migdal:1978az}  occurs when $\mu_I \geq m_\pi$.
Later on, the deconfined superconducting phase has been proposed at high $\mu_I$ and sketched hypothetically on the phase diagram with regard to QCD crossover in terms of $\mu_I$ and temperature~\cite{Son:2000xc}.
Remarkably, unlike $\mu_B$, finite $\mu_I$ does not incur the sign problem so there have been numerous LQCD efforts~\cite{Kogut:2002tm,Kogut:2002zg,Kogut:2004zg,Brandt:2017oyy,Brandt:2022hwy,Abbott:2023coj} 
reinforcing theoretical arguments.
The relevance of isospin asymmetry branches out to astrophysics, where cutting-edge subjects based on the $\pi^\pm$ condensate include the newly proposed compact stars called pion stars~\cite{Brandt:2018bwq,Stashko:2023gnn,Kirichenkov:2023omy}, as well as the pion condensate in early universe with lepton flavor asymmetry signified by gravitational waves~\cite{Vovchenko:2020crk}.

Particularly to our interest is the meanwhile baryon-rich context, e.g., neutron stars. 
Therein the $\mu_I$ is responsible for the nuclear symmetry energy, whose constraining has been actively discussed~\cite{Drischler:2020hwi,PREX:2021umo,Reed:2021nqk,Neill:2022psd}.
The origin of magnetic fields in neutron stars remains an open question.
The answer shall lie in the innermost structure of the neutron star.
A popular proposal is the proton superconductivity that occurs during the cooling via neutrino emission~\cite{1969Natur.224..673B, Graber:2016imq}, which could last over $10^6$ yr~\cite{Ho:2017bia}.
It has been indicated that for magnetic field $B\lesssim 10^{15}$ G the majority of neutron star interior is in the Meissner state expelling magnetic fluxes whilst for stronger $B$ a type-II superconducting phase is possible. 
The latter case could retain magnetic fields inside the neutron star core.
Exploring the generation of magnetic fields in such a context is one of our motivations. 

To this end, we target a low energy hadronic regime under $B$ with both finite $\mu_B$ and $\mu_I$, adopting the method of ChPT and the Skyrme model.
We are interested in phases containing vortices that carry the magnetic flux. 
The vortex formed by pions with finite $\mu_I$ and $B$ but at $\mu_B=0$ has been studied in Refs.~\cite{Adhikari:2015wva,Adhikari:2018fwm,Adhikari:2022cks,Gronli:2022cri}.
In particular, Ref.~\cite{Gronli:2022cri} yields a phase diagram 
of CSL and an Abrikosov vortex lattice (AVL).
In terms of incorporating baryons, Ref.~\cite{Canfora:2020uwf} considers a baryon arising in a pionic crystal, but without effects from $\mu_B$.
On the other hand, the authors in refs.~\cite{Evans:2022hwr,Evans:2023hms} 
have considered an AVL with $\mu_B$ and $B$ whereas neglected $\mu_I$. Such a scenario appears as a fate of the instability of CSL under $\pi^\pm$ condensate at large $B$ and $\mu_B$~\cite{Brauner:2016pko}. 
Distinctively, in the present study we turn on both $\mu_B$ and $\mu_I$, and innovatively introduce a spatially dependent phase angle $\varphi(x)$ of $\pi^0$, on top of the azimuthal $\phi$ of $\pi^\pm$. 
Intuitively, the added $\pi^0$ winding would decrease the vortex energy via the WZW term just as in CSL so our updated vortex configuration could feature lower energy and conquer the ground state.  

In our scenario, the essential difference from aforementioned proposals consists in that our vortex carries a conserved baryon number, {\it i.e.}, the spatial integration of $j_B^0$. Thus our vortex is of baryonic nature, entitled the ``baryonic vortex''.
Yet such an idea has its clue from the findings of Refs.~\cite{Gudnason:2014hsa,Gudnason:2014jga,Gudnason:2016yix,Nitta:2015tua}
that twisting $U(1)$ modulus of a vortex yields a baryon number 
in the ChPT or Skyrme model, 
where Skyrmions are realized as sine-Gordon solitons 
inside a vortex, and such composites
are called vortex-Skyrmions.
Importantly, we couple a baryonic vortex within the Skyrme model to the Maxwell electrodynamics. 
In this way, the magnetic field is solved self-consistently. 
Eventually, by evaluating the energy of the baryonic vortex we determine the critical $\mu_B$, above which the baryonic vortex bears lower energy than the uniform $\pi^\pm$ condensate, under a certain $\mu_I$.
In effect, we stipulate the phase boundary between 
the $\pi^{\pm}$ condensate phase where a vortex could be created but as an excited state under an external magnetic field, 
and 
the baryonic vortex phase where magnetic fluxes are created spontaneously without external magnetic fields.

The present paper is organized as follows. 
We begin with a recap of our theoretical framework in Sec.~\ref{sec: chiral lagrangian}; the $SU(2)$ ChPT coupled with Maxwell electrodynamics and the extension to the Skyrme model.
Next, in Sec.~\ref{sec: vortex ansatz} we lay out our scenario, from the innovative vortex Skyrmion Ansatz tailored for a self-generated axial magnetic field to the expression of the action in terms of such Ansatz. 
Then naturally we present the soliton solution and its properties in Sec.~\ref{sec: soliton configuration}. 
Quintessential results are encompassed in Sec.~\ref{sec: phase diagram}, {\it{i.e.}}, the phase diagram of the $\pi^\pm$ condensate and the baryonic vortex phase parameterized by $\mu_I$ and $\mu_B$.  
In Sec.~\ref{sec: chpt}, we explore our scenario in ChPT without the Skyrme term, commenting on the validity and necessity of our proposal.   
Finally, Sec.~\ref{sec: conclusion} is devoted to the summary of the present study and the outlook for explaining magnetic fields in neutron stars. 
In the appendix, we give details of the equation of motion (EOM) for our Ansatz.

%%%%%%%%%%%%%%%%
\section{Chiral Lagrangian and the Skyrme Model}\label{sec: chiral lagrangian}
In the present work,
the targeted physics regime is the low energy hadronic phase in the chiral limit.
We input a homogeneous isospin chemical potential
$\mu_{I}$ and then presume the ground state is the charged
pion condensation.
The chiral Lagrangian in terms of the pionic $SU\left(2\right)$ field $\Sigma$
reads
\begin{equation}
\mathcal{L}_{\text{chiral}}=\frac{f_{\pi}^{2}}{2}\mathrm{Tr}\left(D_{\mu}\Sigma^{\dagger}D^{\mu}\Sigma\right),
\end{equation}
where $f_{\pi}$ is the pion decay constant and $D_{\mu}$ is a covariant
derivative promoted by the $U\left(1\right)$ electromagnetic (EM) gauge field $A_{\mu}$, together with the gauge field for the third component of isospin $A_{\mu}^I$, {\it i.e.},
\begin{equation}
D_{\mu}\Sigma=\partial_{\mu}\Sigma-i\left(A_{\mu}+A_{\mu}^I\right)\left[Q,\Sigma\right].
\end{equation}
In our notation the elementary charge $e$ is absorbed into the definition
of the gauge fields. The electric charge matrix $Q$ amounts to $\mathbb{I}/6+\tau^{3}/2$ with the Pauli matrices $\tau^{i}$. 
The $\mu_I$ comes into play via $A_{\mu}^I=\delta_{\mu 0}\mu_I$, acting as an electric potential in the covariant derivative, but contributing differently to the gauged Wess-Zumino-Witten (WZW) term:
\begin{equation}
\mathcal{L}_{\text{WZW}}=\left(q A_{\mu}+A_{\mu}^{B}\right)j_{B}^{\mu},
\label{eq:wzwdef}
\end{equation}
which originates from the triangle anomaly. 
In the present case $q=1/2$ and
the topological current $j_{B}^{\mu}$ is known as 
\begin{align}
j_{B}^{\mu} & =\frac{1}{24\pi^{2}}\epsilon^{\mu\nu\alpha\beta}\mathrm{Tr}\left[L_{\nu}L_{\alpha}L_{\beta}-\frac{3iQ}{2}\left(F_{\nu\alpha}+F_{\nu\alpha}^I\right)\left(L_{\beta}-R_{\beta}\right)\right],\nonumber \\
 & \equiv\frac{1}{24\pi^{2}}\epsilon^{\mu\nu\alpha\beta}\mathrm{Tr}\bigg\{ l_{\nu}l_{\alpha}l_{\beta}-3iQ\partial_{\nu}\left[\left(A_{\alpha}+A_{\alpha}^I\right)\left(l_{\beta}-r_{\beta}\right)\right]\bigg\} ,\label{eq:jBdef}
\end{align}
where the $SU\left(2\right)$ currents are $l_{\mu}=\Sigma^{\dagger}\partial_{\mu}\Sigma$
and $r_{\mu}=\Sigma\partial_{\mu}\Sigma^{\dagger}$, whose covariant
versions read $L_{\mu}=\Sigma^{\dagger}D_{\mu}\Sigma$ and $R_{\mu}=\Sigma D_{\mu}\Sigma^{\dagger}$.
Apparently, $F_{\mu\nu}=\partial_{\mu}A_{\nu}-\partial_{\nu}A_{\mu}$ is the field strength of EM gauge field. So forth defined is the isospin counterpart $F_{\mu\nu}^I$.  
In short, though affecting the constitution of $j_B^{\mu}$, $A_{\mu}^I$ is absent in the bracket among Eq.~\eqref{eq:wzwdef} because $\mu_I$ is associated with the $SU\left(2\right)$, rather than $U(1)$, gauging of the WZW term. 

We are to deal with a dynamical magnetic field so the contribution from the self-coupling of EM
fields to the Lagrangian should be included
\begin{equation}
\mathcal{L}_{\text{EM}}=\frac{1}{4}F_{\mu\nu}F^{\mu\nu},
\end{equation}
in which $F_{\mu\nu}=\partial_{\mu}A_{\nu}-\partial_{\nu}A_{\mu}$
is the field strength. 
Keeping in mind the chiral perturbation theory (ChPT) is an effective
theory based on momentum $p$ expansion, we remark that there exist two
kinds of power counting schemes, concerning the gauge field; one
is to count $e\sim\mathcal{O}\left(p\right)$, and the other is to count
$A_{\mu}\sim\mathcal{O}\left(p\right)$. The bifurcation would influence
the order of $\mathcal{L}_{\mathrm{EM}}$ which actually contains
the abbreviated $1/e^{2}$ as a factor. In the former scheme $\mathcal{L}_{\mathrm{EM}}\sim\mathcal{O}\left(p^{2}\right)$
so higher order terms in the Lagrangian typified by the Skyrme term
could be regarded as the next leading order while in the latter
scheme, $\mathcal{L}_{\mathrm{EM}}\sim\mathcal{O}\left(p^{4}\right)$
makes it necessary to go beyond the $\mathcal{O}(p^2)$ chiral Lagrangian to evaluate EM fields consistently. 
Moreover,
one can check that a similar nuance of power counting applies to the $\mathcal{L}_{\text{WZW}}$. We would further address the consequence brought by such a nuance in Sec.~\ref{sec: chpt}. 
Our choice of the
$\mathcal{O}\left(p^{4}\right)$ term would be the Skyrme term
\begin{equation}
\mathcal{L}_{\text{Skyrme}}=\frac{1}{32s^{2}}\mathrm{Tr}\left(\left[L_{\mu},L_{\nu}\right]\left[L^{\mu},L^{\nu}\right]\right).
\end{equation}
The dimensionless parameter $s$ was originally fixed by matching
the experiments and would be taken as a constant in our study.

\section{Vortex Skyrmion Ansatz}\label{sec: vortex ansatz}

For simplicity, we specify our scenario to be a static vortex line
in a longitudinal magnetic field, which means that we shut down time dependence
and retain a cylindrical symmetry in the space. The vortex Skyrmion Ansatz
is written as 

\begin{equation}
\Sigma=\sigma+i\tau^{i}\pi^{i}=\left(\begin{array}{cc}
e^{i\varphi}\cos\alpha & e^{-i\phi}\sin\alpha\\
e^{i\phi}\sin\alpha & e^{-i\varphi}\cos\alpha
\end{array}\right).
\label{eq:ansatz}
\end{equation}
The $\phi$ is the azimuthal angle among the cylindrical coordinate
system $\left(\rho,\phi,z\right)$. $\alpha$ and $\varphi$
are functions of $\rho$ and $z$, which are to be solved. To produce
the magnetic field $\boldsymbol{B}=B\left(\rho\right)\hat{z}$, the needed
spatial component of the gauge field is $\boldsymbol{A}=A_{\phi}\left(\rho\right)\hat{\phi}$.
Meanwhile, we remind the temporal component $A_{0}=\mu_{I}$ being
the isospin chemical potential. We would further simply our setup
by presuming $\varphi=\varphi\left(z\right)$ and $\alpha\left(\rho\right)$.
We remind that our Ansatz does NOT guarantee the configuration with
the lowest energy, but represents generally an excited state on top of the $\pi^{\pm}$ condensate background. Hence at spatial infinity,
the Ansatz should recover the state of the superconducting $\pi^{\pm}$
ring with vanishing $\pi^{0}$, prescribing the boundary condition:

\begin{equation}
\alpha\left(0\right)=0,\quad\alpha\left(\infty\right)=\frac{\pi}{2}.\label{eq:bcalpha}
\end{equation}
The condition at origin is instead a pure $\pi^0$ sector winding with $\varphi(z)$, which is reminiscent to CSL.
%requested by avoiding singularity and contributing to the winding number.

%The ground state could be altered 
Intuitively, when there is a finite baryon chemical potential
$\mu_{B}$, such a nontrivial $\pi^{0}$ winding could contribute to a topologically preserved baryon number which
is absent in $\pi^{\pm}$ condensate. 
The energy could therefore be reduced by
the coupling of the baryon number to $\mu_{B}$ through the WZW term.
If such a reduction is large enough, we conjecture
that our vortex Skyrmion could be of lower energy than the $\pi^{\pm}$ condensate.
This is our original motivation to consider the vortex Skyrmion Ansatz~\eqref{eq:ansatz}.
%Along this thread, we think of our vortex ansatz with a nontrivial winding of $\varphi\left(z\right)$. 
To this end, we scrutinize the constitution
of the baryon density in terms of our Ansatz:
\begin{align}
j_{B}^{0} & =\frac{1}{4\pi^{2}}\left[\left(\frac{1}{\rho}+A_{\phi}\right)\sin2\alpha\partial_{\rho}\alpha\partial_{z}\varphi-\frac{1}{\rho}\frac{\partial}{\partial\rho}\left(\rho A_{\phi}\right)\cos^{2}\alpha\partial_{z}\varphi\right]\nonumber \\
 & =\frac{1}{4\pi^{2}}\left[\frac{1}{\rho}\sin2\alpha\partial_{\rho}\alpha\partial_{z}\varphi-\frac{1}{\rho}\frac{\partial}{\partial\rho}\left(\rho A_{\phi}\cos^{2}\alpha\right)\partial_{z}\varphi\right].\label{eq:jB}
\end{align}
Notably, the second portion in the bracket, a surface term after the
spatial integral, vanish in our scenario given $\alpha\left(\infty\right)=\pi/2$.
This is in sheer contrast to the often discussed $\pi^{0}$ domain
walls, or the CSL, in which the surface term yields the familiar
$-\boldsymbol{B}\cdot\nabla\varphi/4\pi^{2}$ whilst the first term
in the bracket of Eq.~\eqref{eq:jB} vanishes due to the absence of
charged $\pi^{\pm}$.
%seen from $\alpha\equiv0$
Evaluating the topological
baryon charge stipulates the boundary conditions for $\varphi\left(z\right)$,
{\it i.e.}, 
\begin{align}
\int d^{3}xj_{B}^{0} & =\frac{1}{4\pi^{2}}\int\rho d\rho d\phi dz\frac{1}{\rho}\sin2\alpha\partial_{\rho}\alpha\partial_{z}\varphi=\frac{N_B}{2\pi}\varphi\left(z\right)\big|_{z=0}^{z=L},\label{eq:j0}
\end{align}
where we have used the boundary condition of $\alpha\left(\rho\right)$
in Eq.~\eqref{eq:bcalpha}.
Naturally, from the longitudinal periodicity 
we set one baryon in each period $L$ and denote the total baryon number as $N_B$.
Then the boundary condition can be written as: 
%of $\varphi\left(z\right)$:
\begin{equation}
\varphi\left(0\right)=0,\quad\varphi\left(L\right)=2\pi,\label{eq:bcphi}
\end{equation}
so that $\int d^{3}xj_{B}^{0}\equiv N_B$. 
From here one can clearly see such a vortex Skyrmion 
%is composed of not only $\pi^{\pm}$ but also $\pi^{0}$ and 
carries a conserved baryon number. 
This is the connotation that we call our vortex Skyrmion a ``baryonic vortex''.
Under such a setup, we note apart
from $j_{B}^{0}$, the baryon current features nontrivial azimuthal
component relying on the $\mu_{I}$:
\begin{equation}
j_{B}^{\phi}=\frac{1}{4\pi^{2}}\frac{\partial}{\partial\rho}\left(\mu_{I}\cos^{2}\alpha\right)\partial_{z}\varphi.
\end{equation}
Together with Eq.~\eqref{eq:jB} it leads to the following WZW term:
\begin{align}
\mathcal{L}_{\text{WZW}} & =\mu_{B}j_{B}^{0}+qA_{\phi}j_{B}^{\phi}.\label{eq:wzw}
\end{align}
We clarify that throughout this work, as a first step, we consider
homogenous $\mu_{I}$ and $\mu_{B}$ for simplicity. %We can already see the different way they influence our system. The $\mu_{B}$ related term contributes to the energy via a plainly integrated $-\mu_{B}$ for a single vortex, irrelevant to the equation of motion (EOM). However, $\mu_{I}$ enters into the EOMs of both $\alpha$ and $A_{\phi}$, impacting the configurations of the vortex and the magnetic field.

To concretely solve these configurations, we need the last boundary
condition for $A_{\phi}.$ It can be defined in view of the chiral
Lagrangian, whose specific expression reads: 
\begin{equation}
\mathcal{L}_{\text{chiral}}=-\frac{f_{\pi}^{2}}{2}\left\{ \left(\partial_{\rho}\alpha\right)^{2}+\left[\left(\frac{1}{\rho}+A_{\phi}\right)^{2}-\mu_{I}^{2}\right]\sin^{2}\alpha+\cos^{2}\alpha\left(\partial_{z}\varphi\right)^{2}\right\} .
\end{equation}
%In the superconducting region
At $\rho\rightarrow\infty$, the kinetic
energy $-\mathcal{L}_{\text{chiral}}$ density needs to vanish (except for a constant level brought
by $\mu_{I}^{2}$ term), which requires $A_{\phi}=-1/\rho$, a pure
gauge field. For numerical convenience we define $a\left(\rho\right)\equiv\rho A_{\phi}\left(\rho\right)$
and dictate its boundary condition as follows
\begin{equation}
a\left(0\right)=0,\quad a\left(\infty\right)=-1.\label{eq:bca}
\end{equation}
For completeness, we also display the expression of the Skyrme term
under our Ansatz
\begin{align}
\mathcal{L}_{\text{Skyrme}}= & -\frac{1}{2s^{2}}\left\{ \left(\partial_{\rho}\alpha\right)^{2}\cos^{2}\alpha\left(\partial_{z}\varphi\right)^{2}\right.\nonumber \\
 & \left.+\left[\left(\frac{1}{\rho}+A_{\phi}\right)^{2}-\mu_{I}^{2}\right]\sin^{2}\alpha\left[\left(\partial_{\rho}\alpha\right)^{2}+\cos^{2}\alpha\left(\partial_{z}\varphi\right)^{2}\right]\right\} .
\end{align}
As we will show soon, the Skyrme term plays the role to guarantee
a finite size of the soliton, overcoming the Derrick's scaling law,
as it did in the Skyrme model description of a baryon.

In summary, what we need to solve in the current study are the functions
$\alpha\left(\rho\right)$, $a\left(\rho\right)$ and $\varphi\left(z\right)$
under boundary conditions Eqs.~\eqref{eq:bcalpha}, \eqref{eq:bcphi}
and \eqref{eq:bca}.
As long as we do not consider the time evolution, solving the EOM is tantamount to finding the soliton configuration that minimizes
the energy functional corresponding to $\mathcal{L}=\mathcal{L}_{\text{chiral}}+\mathcal{L}_{\text{Skyrme}}+\mathcal{L}_{\text{WZW}}+\mathcal{L}_{\text{EM}}$, specifically
\begin{equation}
H=\int d^{3}x\mathcal{H}=\int2\pi\rho d\rho dz\left[\frac{1}{2}\left(\frac{1}{\rho}\partial_{\rho}a\right)^{2}-\mathcal{L}_{\text{chiral}}-\mathcal{L}_{\text{Skyrme}}-\mathcal{L}_{\text{WZW}}\right].\label{eq:H}
\end{equation}
In practice, we would employ the variational principle $\delta H=0$
to solve the system. 
We remark that the
variational principle with respect to $A_{\mu}$ is equivalent to
the Maxwell equation $\partial_{\mu}F^{\mu\nu}=j_{Q}^{\nu}$ with
the electric current $j_{Q}^{\mu}$ defined by 
\begin{equation}
j_{Q}^{\mu}=qj_{B}^{\mu}+j_{I}^{\mu},
\end{equation}
among which the non-anomalous part is dubbed the isospin current 
\begin{align}
j_{I}^{\mu}\equiv & \frac{\delta}{\delta A_{\mu}}\left(\mathcal{L}_{\text{chiral}}+\mathcal{L}_{\text{Skyrme}}\right)\nonumber \\
= & \delta^{\mu\phi}\left(\frac{1}{\rho}+A_{\phi}\right)\sin^{2}\alpha\left\{ f_{\pi}^{2}+\frac{1}{s^{2}}\left[\left(\partial_{\rho}\alpha\right)^{2}+\left(\partial_{z}\varphi\right)^{2}\cos^{2}\alpha\right]\right\} .
\end{align}
The specific
form of the EOMs, two coupled second order ordinary differential equations, is clarified in the Appendix.

It is worth mentioning that the pion mass term, if added, would impact not only
the Lagrangian by $\mathcal{L}_{\text{mass}}=(1/2)f_{\pi}^{2}m_{\pi}^{2}\mathrm{ReTr}\Sigma=f_{\pi}^{2}m_{\pi}^{2}\cos\varphi\cos\alpha,$
but also the validity of the above boundary conditions Eqs.~\eqref{eq:bcalpha} and \eqref{eq:bcphi}.
That is, when $\mu_{I}>m_{\pi}$ the $\pi^{\pm}$
condensate coexists with a finite $\pi^{0}$
density on the spatial boundary; 
%{\it i.e.}, in our ansatz 
$\alpha\left(\rho\rightarrow\infty\right)\rightarrow\cos^{-1}\left(m_{\pi}^{2}/\mu_{I}^{2}\right)$.
Moreover, in this case $\varphi$ would necessarily feature nontrivial dependence
on $\rho$, invalidating our simplified $\varphi=\varphi(z)$. 
The result should contain richer topological structures, which
would be presented shortly in a followup work.
Nevertheless, as we know the $\pi^\pm$ condensate occurs only when $\mu_I > m_\pi$, we would tune the input $\mu_I$ from $m_\pi$, rather than zero, to larger. So that the compatibility of the simplified boundary conditions is validated.

%%%%%%%%%%%%%%%%%%
\section{Baryonic Vortex  Configurations}\label{sec: soliton configuration}

We solve the vortex Skyrmion (our ansatz for the baryonic vortex) numerically. 
For such purpose, physical
quantities are rescaled to be dimensionless, e.g.,
\begin{equation}
\tilde{\rho}\equiv\rho f_{\pi}s,\quad\tilde{z}\equiv zf_{\pi}s,\quad\tilde{A_{\mu}}\equiv\frac{A_{\mu}}{f_{\pi}s},\quad\tilde{H}\equiv\frac{H}{f_{\pi}s^{-1}}.\label{eq:rescale}
\end{equation}
Eventually, we will present our results with units
made of $f_{\pi}=54\text{ MeV}$ and $s=5.45$, whose specific values
are taken from earlier literature \cite{Adkins:1983ya}\footnote{If we include pion mass $m_{\pi}=138\text{ MeV}$, we could use $s=4.84$}. Moreover, as we clarified though we would consider the chiral limit, our input $\mu_I$ would start from the effective $m_\pi=139 \text{ MeV}$, which reads $\Tilde{\mu_I}>0.47$ via our dimensionless unit. With these conventions specified, we proceed to preset the vortex Skyrmion solutions.

\begin{figure}[!htb] 
\minipage{0.46\textwidth}   
\includegraphics[width=\linewidth]{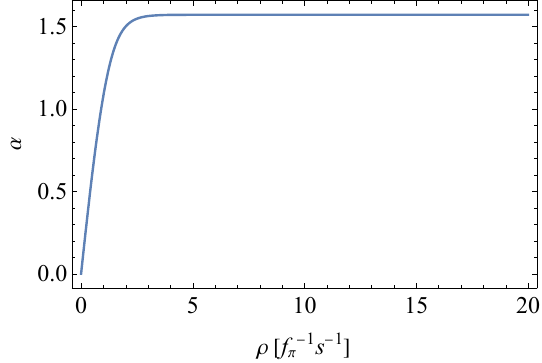}   
%\caption*{}   
%\label{fig:awesome_image1} 
\endminipage\hfill 
\minipage{0.5\textwidth}   
\includegraphics[width=\linewidth]{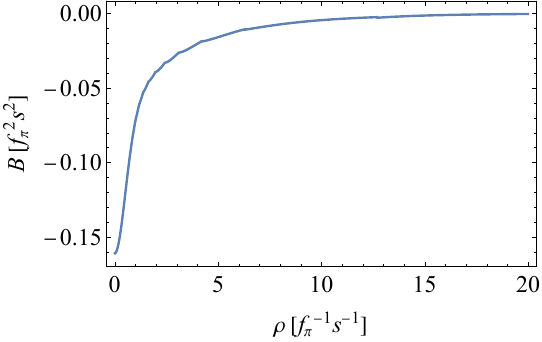}   
%\caption*{}   
%\label{fig:awesome_image1} 
\endminipage\hfill 
\caption{Vortex Skyrmion (left) and magnetic field (right) at $\mu_I=0.5 f_\pi s$ and $k=2.0 f_\pi s$.}
\label{fig:profile}
\end{figure}

One can immediately observe that the $z$-dependence is separated due to our Ansatz. 
The EOM of $\varphi$ is nothing but $\partial_{z}^{2}\varphi=0$,
leading to 
\begin{equation}
\varphi=kz;\quad k\equiv\frac{2\pi}{L}.
\label{eq:varphi}
\end{equation}
Here the $k$ marks a typical momentum scale of the system. %which should be comparable to our input $\mu_{I}$ and $\mu_{B}$.
Obviously, $\mu_{B}$ would
not enter into the transverse EOMs. 
Then the profile ($\rho$-dependence) of the baryonic vortex is up to $\mu_I$ and $k$.
We pick one fixed 
$\mu_{I}=0.5f_{\pi}s$ to delineate the profile and illustrate
how its properties depend on the longitudinal wave vector $k$. 
In Fig.~\ref{fig:profile},
we show the exemplary profile where $B\left(\rho\right)=\partial_{\rho}a\left(\rho\right)/\rho$.

\begin{figure}[!htb] 
\centering       
\includegraphics[width=0.6\columnwidth]{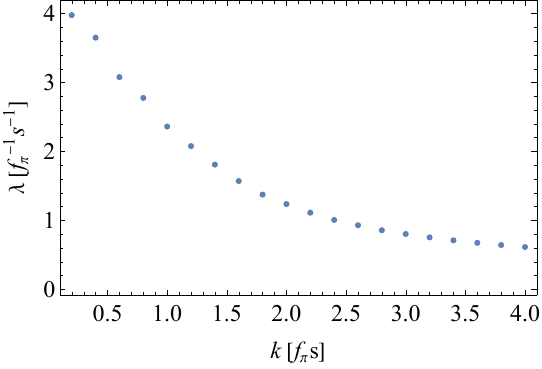}   
%\caption*{}   
%\label{fig:awesome_image1} 
%\endminipage\hfill 
\caption{Half $1/e$ width of the vortex at $\mu_I=0.5 f_\pi s$.}
\label{fig:property}
\end{figure}

It has been a consensus that at an asymptotic distance, the solution is
approaching the $\pi^{\pm}$ condensate. 
Therein the magnetic field
should behave as in a superconductor. 
Then one important physical
scale is the penetration depth of the magnetic field. 
Rigorously speaking,
for a general range of $\rho$, the $B\left(\rho\right)$ in our scenario
is not subject to the London equation. 
Thus the penetration depth
applies only in asymptotic regions. 
But in such a spirit, based on the wave packet shape of $B\left(\rho\right)$ seen in Fig.~\ref{fig:profile} (right),
one can always define the ``$1/e$ width'' which amounts to $2\lambda$
with $B\left(\lambda\right)=B_{0}/e$, to characterize the transverse scale of
the vortex. We exhibit in Fig.~\ref{fig:property} (right) the $\lambda$
at different $k$, demonstrating the positive correlation between
longitudinal and transverse sizes of the vortex. Such a correlation
could be understood intuitively way by quoting the concept of London penetration
depth $\lambda_{L}$ after the asymptotic expansion
\begin{align}
\lambda_{L}^{-1} & \sim m_{A}\equiv\sin\alpha\sqrt{f_{\pi}^{2}+\frac{1}{s^{2}}\left[\left(\partial_{\rho}\alpha\right)^{2}+k^{2}\cos^{2}\alpha\right]},\label{eq:lambda}
\end{align}
where $m_{A}$ is the effective mass of the gauge field which can
be read off from the quadratic term of $A_{\phi}$ among the energy
functional.
At spatial infinity, $\lambda_{L}$ reaches its known result for the
$\pi^{\pm}$ condensate within ChPT framework $\lambda_{L}\left(\rho\rightarrow\infty\right)\rightarrow f_{\pi}^{-1}$~\cite{Adhikari:2015wva}.
Here the heuristic generalization of Eq.~\eqref{eq:lambda} is intended
to show $\lambda_{L}$ could receive corrections which are positively correlated
with $L=2\pi/k$, and so should $\lambda$. 
This analysis reasonably
explains qualitative features of Fig.~\ref{fig:property} (right).
The correlation between $\lambda$ and $k$ is essential for the Skyrme
term to maintain at $\mathcal{O}\left(p^{4}\right)$ in Lagrangian
density so it competes with lower order terms to guarantee a finite
size soliton.

So far to illustrate the profile we keep $k$ tunable. It needs to be determined by minimizing the total energy where the integration boils down to
\begin{equation}
H=\int d^3x \mathcal{H} = 2\pi N_B \int^L_0 dz \int^{R\rightarrow \infty}_0 \rho d \rho \mathcal{H}.
\end{equation}
We shall
point out a direct spatial integral of the kinetic energy $-\mathcal{L}_{\text{chiral}}$ would diverge.
Nevertheless, the divergent part can be easily subtracted as the level
of the potential energy set by the $\text{\ensuremath{\pi^{\pm}}}$
condensate background {\it i.e.}, 
\begin{equation}
H_{0}\equiv\int d^{3}x\left(-\frac{1}{2}\mu_{I}^{2}\right)=-\frac{1}{2}\mu_{I}^{2}\pi R^{2} N_B L\bigg|_{R\rightarrow\infty},
\end{equation}
which is derived by taking $\alpha=\pi/2$ in $-\mathcal{L}_{\text{chiral}}$. $R\rightarrow\infty$ of the transverse boundary in our current study
describes a single vortex line while $R$ could be finite in our future study of a vortex lattice. 
For now, we treat
$R$ properly as a large cutoff on the numerical stage. Consequently,
we do find the finite $E=H-H_{0}$, whose physical meaning is
the energy cost to create a vortex on top of the homogeneous $\pi^{\pm}$
condensate.  
Whether the baryonic vortex could emerge as a stable state relies on the dependence of $E$ on $k$.
In a thermodynamic perspective, we are employing the grand
canonical ensemble to study a system with fixed total volume $V=\pi R^{2} L N_B$. 
Given that $R$ is an irrelevant constant, minimizing $E$ is equivalent to minimizing the string tension of the system:
\begin{equation}
T=\frac{E}{V} \pi R^2,
\end{equation}
which is a function of only $k$ (for fixed $\mu_{I,B}$). In this way we have substituted the variable $N_B$. The stable (meta-stable)
vortex is reached at a global (local) minimum of $T(k)$, if exists, \textit{i.e.,} 
\begin{equation}
k_{v}=\mathrm{argmin}\; T\left(k\right).
\end{equation}
Below, we will further discover that depending on the value of $\mu_{B}$, such a stable vortex does exist as either an excited state or even a ground state.

\section{Phase Diagram 
with Isospin and Baryon Chemical Potentials}\label{sec: phase diagram}

\begin{figure}[!htb] 
\minipage{0.33\textwidth}   
\includegraphics[width=\linewidth]{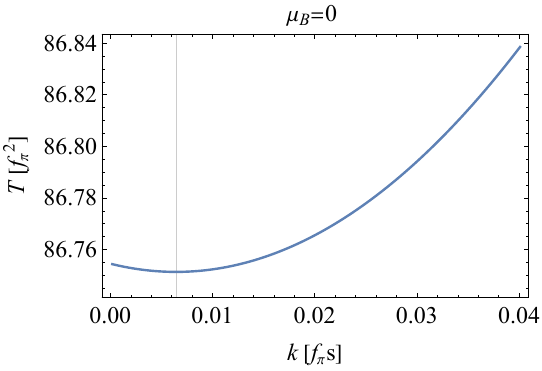}   
%\caption*{}   
%\label{fig:awesome_image1} 
\endminipage\hfill 
\minipage{0.33\textwidth}   
\includegraphics[width=\linewidth]{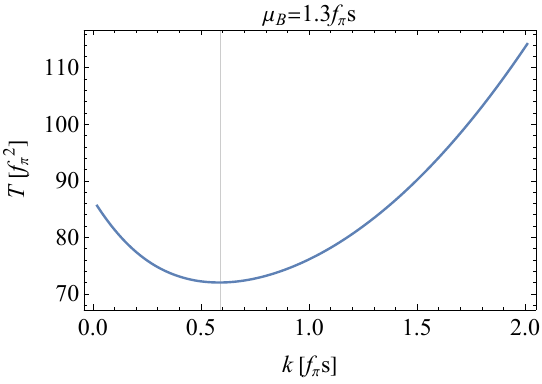}  
%\caption*{}   
%\label{fig:awesome_image1} 
\endminipage\hfill 
\minipage{0.33\textwidth}   
\includegraphics[width=\linewidth]{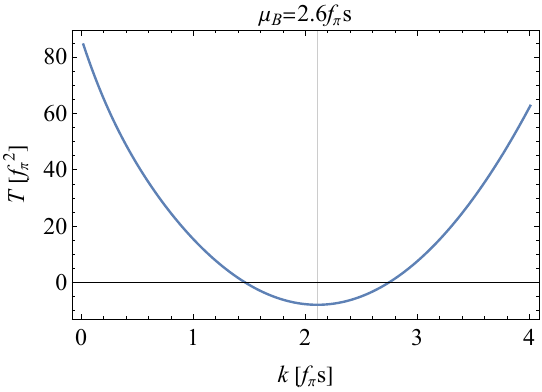}  
%\caption*{}   
%\label{fig:awesome_image1} 
\endminipage\hfill
\caption{String tension for $\mu_I=0.5 f_\pi s$, $\mu_B$ amounts to $0$ (left), $1.3f_\pi s$ (middle) and $2.6f_\pi s$ (right). The grey vertical lines pinpoint the minima $T(k_v)$.}
\label{fig:tension}
\end{figure}

Succeeding to the last section, we adopt the same exemplary $\mu_{I}=0.5f_{\pi}s$
and exhibit the $T(k)$
at different $\mu_{B}$. Remarkably, as shown in
Fig.~\ref{fig:tension} (left), even when $\mu_{B}=0$, there is a
global minimum of $T\left(k\right)$ with $k \neq 0$, which amounts to $86.75f_{\pi}^{2}$
at $k_{v}=0.0065 f_\pi s$. In other words, such a vortex carries energy $2\pi T\left(k_{v}\right)/k_{v}=831\text{ GeV}$ per baryon number,
which is over a hundred times of the heaviest baryon mass, far from
a ground state. 
However, when the $\mu_{B}$ is enhanced, referring to the typical
case exhibited in Fig.~\ref{fig:tension} (middle), we witness a decreasing $T\left(k_{v}\right)$ with an increasing
vortex momentum $k_{v}$.
Importantly, when $\mu_{B}$
is large enough, specifically $\mu_{B}\gtrsim2.518f_{\pi}s\equiv\mu_{B}^{c}$,
the global minimum becomes negative, as shown in \ref{fig:tension} (right), which signifies a new ground state replacing
the homogeneous $\pi^{\pm}$ condensate.
This case is crucially different from the case of $\mu<\mu_B^c$,
in which a positive $T(k_v)$ means that baryonic vortices remain excited states and 
require external magnetic fields to arise, 
as in conventional type-II superconductors.
By contrast, when 
$\mu_{B} \geq \mu_{B}^{c}$, 
baryonic vortices are spontaneously created 
without applying magnetic fields, which implies the spontaneous generation of magnetic fields.

\begin{figure}[!htb] 
\minipage{0.5\textwidth}   
\includegraphics[width=\linewidth]{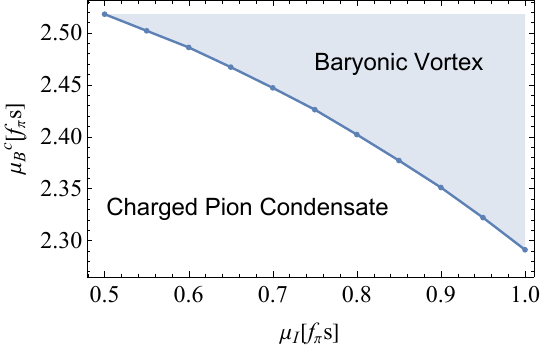}   
%\caption*{}   
%\label{fig:awesome_image1} 
\endminipage\hfill 
\minipage{0.5\textwidth}   
\includegraphics[width=\linewidth]{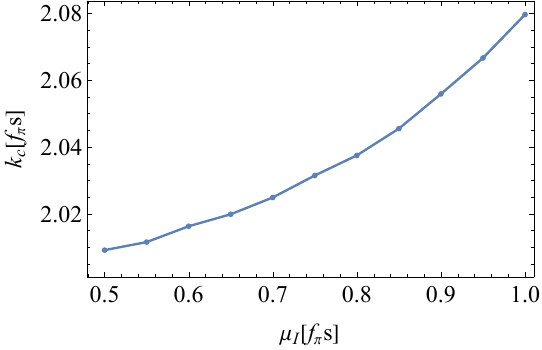}   
%\caption*{}   
%\label{fig:awesome_image1} 
\endminipage\hfill 
\caption{Phase diagram parameterized by the isospin chemical potential $\mu_I$ and baryon chemical potential $\mu_B$ (left),
where the blue dotted line is the critical  $\mu_{B}^c\left(\mu_{I}\right)$ that separates the baryonic vortex phase in blue shade ($\mu_{B}>\mu_{B}^{c}$) and the $\pi^{\pm}$ condensate in the blank ($\mu_{B}<\mu_{B}^{c}$).
Together plotted is the corresponding longitudinal density $k_c$ at each $\mu_I$ (right).}
\label{fig:phase}
\end{figure}

There is such a critical baryon chemical potential
$\mu_{B}^{c}$ for each value of $\mu_{I}$ above $m_\pi$.
The $\mu_{B}^{c}$ as the function of $\mu_{I}$ spans a phase diagram of the two ground states:
baryonic vortex and the $\pi^{\pm}$ condensate.
Such a phase diagram is laid out in Fig.~\ref{fig:phase} 
 (left). 
%For clarity we reiterate that for $\mu_{B}>\mu_{B}^{c}(\mu_{I})$ the baryonic vortex takes up the ground state while for $\mu_{B}<\mu_{B}^{c}(\mu_{I})$ the ground state is $\pi^\pm$ condensate. 
There is a vortex longitudinal density $k_c \equiv k_v(\mu_I,\mu_B^c)$ uniquely determined by the input $\mu_I$ and the corresponding $\mu_B^c$. In Fig.~\ref{fig:phase} (right), we show $k_c$ tends to blow up for large $\mu_I$. 
A related comment is that though $\mu_{B}^{c}(\mu_I)$
seems to go all the way down with an increasing $\mu_{I}$, the deduction
that $\mu_{B}^{c}$ drops to zero for large enough $\mu_{I}$ 
cannot be audaciously made. 
The reason is that $k_{c}$ could severely exceed
its valid range dictated by the ChPT scheme.
Specifically, the nature of momentum expansion stipulates
$k\ll4\pi f_{\pi}$ which is converted to $2.31$ with our momentum
unit $f_{\pi}s$. 
Thus even within the current parameter region of
$\mu_{I}$ and $\mu_{B}$, the $k_{c}$ applies only marginally.
We expect the feasibility of our study to be improved after taking into account fully the pion mass in the dynamics, which would be soon updated in our sequential work.

Despite the caveat, here our simplified analysis already allows us to demonstrate for the first time that in finite isospin QCD, a baryonic vortex with extra $\pi^0$ winding on top of a $\pi^\pm$ condensate could be of lower energy than the homogeneous condensate. Remarkably, such a phenomenon occurs at a reasonable critical baryon chemical potential bounded from above by $\mu_B^c (\mu_I=m_\pi)\sim 743 \text{ MeV}$, which resides within the regime of nuclear matter. 
In other words, in the density window between $\mu_B^c$ and nucleon masses around $940 \text{ MeV}$, a finite $\mu_I$ could yield a ground state made of the baryonic vortex we proposed. In this way, the self-generated magnetic field could be sustained in nucleon-rich contexts with isospin asymmetry, especially neutron stars.   
To further estimate the strength of such magnetic fields, we need to generalize the current scenario to a baryonic vortex lattice so that we could attain the transverse density of the magnetic flux carried by the vortex. 

To this end, Refs.~\cite{Evans:2022hwr,Evans:2023hms} are inspiring yet distinguished from our study in that their field
profile is solved based on a perturbative expansion near the critical
point of the superconducting phase within ChPT, while ours is from
the Skyrme model with no approximation. 
Our baryonic vortex preexists the determination
of the critical magnetic field, allowing for a spontaneous generation of the magnetic field.
According to our power counting scheme,
the Skyrme term is essential for the global minimum of string tension to exist with a finite $k_{v}$.  
We will further substantiate this point in the next section. 

%%%%%%%%%%%%%%%%%%%%%%%%%%%%%
\section{Remark on $\mathcal{O}(p^2)$ Results}\label{sec: chpt}

\begin{figure}[!htb] 
\minipage{0.33\textwidth}   
\includegraphics[width=\linewidth]{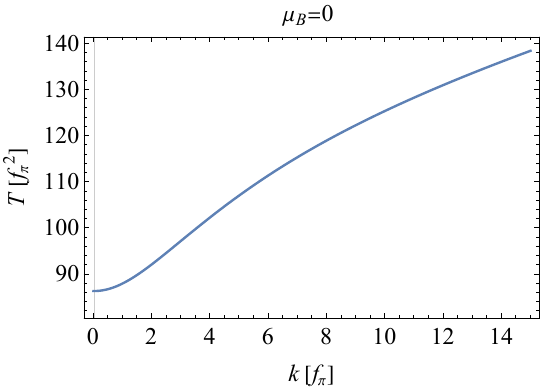}   
%\caption*{}   
%\label{fig:awesome_image1} 
\endminipage\hfill 
\minipage{0.33\textwidth}   
\includegraphics[width=\linewidth]{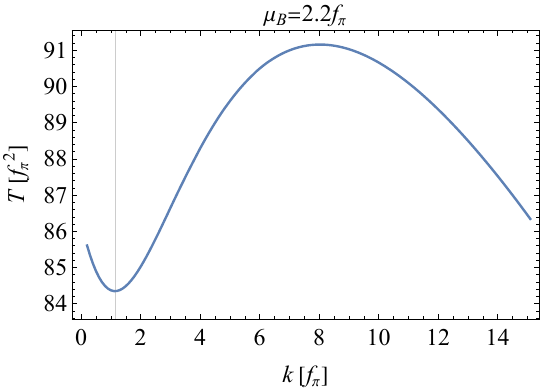}   
%\caption*{}   
%\label{fig:awesome_image1} 
\endminipage\hfill 
\minipage{0.33\textwidth}   
\includegraphics[width=\linewidth]{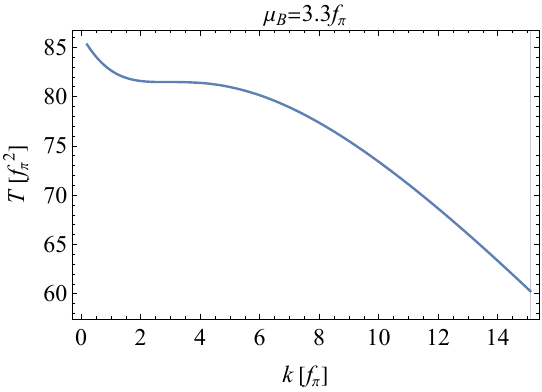}   
%\caption*{}   
%\label{fig:awesome_image1} 
\endminipage\hfill
\caption{String tension evaluated from ChPT without the Skyrme term for $\mu_I=0.5 f_\pi s$ while $\mu_B$ amounts to $0$ (left), $2.2f_\pi$ (middle) and $3.3f_\pi$ (right).}
\label{fig:tensionchpt}
\end{figure}

To unveil the role played by the Skyrme term and the valid range of our proposal, we compare our scenario with vortices in ${\cal O}(p^2)$
ChPT without the Skyrme term. 
In this case, the rescaling in Eq.~\eqref{eq:rescale} shall be modified, with momentum in unit $f_{\pi}$. A counterpart of Fig.~\ref{fig:tension} is exhibited
in Fig.~\ref{fig:tensionchpt}, where we maintain the common input
$\mu_{I}=0.5\times5.45f_{\pi}$. Comparing Fig.~\ref{fig:tensionchpt} (left) and Fig.~\ref{fig:tension} (left), we observe that for $\mu_{B}=0$,
the behavior of $T\left(k\right)$ is similar between ${\cal O}(p^2)$
ChPT and the Skyrme
model. 
It features a global minimum where $k_{v}$ with and without the Skyrme term are $0.0065f_{\pi}s$ and
$0.042f_{\pi}$ respectively, close to each other. 
The $T\left(k_{v}\right)$
are almost the same as well; $86.75f_{\pi}^{2}$ with the Skyrme term,
and $86.25f_{\pi}^{2}$ without it. 
So we affirm that for the previous
studies 
\cite{Adhikari:2015wva,Adhikari:2018fwm,Adhikari:2022cks,Gronli:2022cri} without $\mu_{B}$, ${\cal O}(p^2)$ ChPT used by them is sufficient and the Skyrme term in our proposal would only
produce minor corrections on their results. 
Moreover, their scenarios correspond to the $k=0$ case in our study. 
As one can see from either Fig.~\ref{fig:tension} or Fig.~\ref{fig:tensionchpt}, the well of $T(k)$ is very shallow.
Thus although our newly introduced finite $k_v$ configuration lowers the minimum vortex energy, such an effect is not prominent for the case with only $\mu_I$. 
However, the situation is drastically different for $\mu_B \neq 0$.

We highlight that once a finite $\mu_{B}$
is turned on, Fig.~\ref{fig:tensionchpt} (middle) exemplifies
the eccentric behavior of $T(k)$: the minimum becomes
local and $T\left(k\right)$ decreases with increasing $k$ when $k$
is large. Such an ultraviolet behavior seems to indicate 
the instability of the system;
a runaway ground
state with infinitely large baryon density. This indication is even
more pronounced when $\mu_{B}$ is larger, as shown in Fig.~\ref{fig:tensionchpt} (right).
The local minimum of $T\left(k\right)$ disappears when  $\mu_{B}\gtrsim3.3f_{\pi}$ and
the $T\left(k\right)$ becomes a monotonically decreasing function. 
We regard such an infinitely 
dense vortex as an artifact brought by the the breakdown of ChPT expansion hierarchy. 
The negative contribution to $T(k)$ is from the $\mathcal{O}\left(p^{3}\right)$ WZW term~\eqref{eq:wzw} which leads to a factor $-\mu_{B}k$ after the spatial integration. 
For ultraviolet $k$, such negative contribution would always surpass the positive contribution from the $\mathcal{O}\left(p^{2}\right)$ chiral Lagrangian. 
This is exactly why we must add the Skyrme term, an $\mathcal{O}\left(p^{4}\right)$
positive contribution, for a stable baryonic vortex of finite longitudinal density.
Then once the vortex with finite $k_v$ stands as a solution, it carries the baryon number, leading to the WZW coupling. This is the physical mechanism of a negative string tension when $\mu_B$ is large, which is exclusive to our proposal that goes beyond $\mathcal{O}(p^2)$ order.

\section{Conclusion}\label{sec: conclusion}
In this study, we have proposed a baryonic vortex phase as a novel phase for dense hadronic matter with finite $\mu_B$ and $\mu_I$. 
We determine the boundary $\mu_B=\mu_B^c(\mu_I)$ between such a phase and the charged pion $\pi^\pm$ condensate in the phase diagram in Fig.~\ref{fig:phase}.
%We also elucidate that the ChPT framework applies to a limited magnitude of $\mu_I$, requiring the remedy from higher order derivative terms such as the Skyrme term.
In the presumed regime 
$\mu_I \gtrsim m_\pi$, 
the critical $\mu_B^c$ starts from around 743 MeV and decreases with an increasing $\mu_I$. When $\mu_B > \mu_B^c$ the baryonic vortex is energetically preferred over the homogeneous $\pi^\pm$ condensate. 

Vortices emerging from $\pi^\pm$ condensate is not a new story.
They form AVL under an external magnetic field, similar to metallic type-II superconductors,
as studied in the previous works 
\cite{Adhikari:2015wva,Adhikari:2018fwm,Adhikari:2022cks,Gronli:2022cri}.
In such cases, the optimal lattice spacing is determined by the strength of the applied magnetic field, given magnetic flux is quantized for each vortex.
The detail of the interaction between vortices is unimportant as far as the vortices repel each other.
Our results for $0\leq \mu_B<\mu_B^c$ update these works in the sense of lowering the vortex energy via a nontrivial WZW coupling of the baryon current to the $\mu_B$, which means a less demanding applied magnetic field to excite the vortices from $\pi^\pm$ condensate. 

In contrast, our scenario of the baryonic vortex phase at $\mu_B>\mu_B^c$ is completely original, in which a ``baryonic vortex lattice (BVL)'' is expected to emerge even with NO applied magnetic field. 
In such a circumstance, the self-generated magnetic field, according to our evaluation, could reach $10^{-1}f_{\pi}^{2}s^{2}\sim10^{17}\text{ Gauss}$ near the vortex core.  
Nevertheless, the magnitude of the macroscopic magnetic field is still to be quantified, which hinges on the lattice spacing, or the baryon density of the BVL.  
Unlike in the AVL,
the lowest energy configuration of a BVL must be achieved from the balance between the negative string tension of each vortex and the repulsion between vortices.
Hence the detail of the interaction does matter. This part
would be figured out in our subsequent work, which shall illuminate the modification on the critical magnetic field needed when $\mu_B<\mu_B^c$, and the quantification of baryon density together with the spontaneously generated magnetic field when $\mu_B>\mu_B^c$.   

The significance of such researches is twofold.
The first is the relevance in astrophysics.
We conceive that the self-generated and conserved magnetic flux of our baryonic vortex possibly explains
the origin of the long lived ($>10^6$ Yrs) magnetic field inside magnetars. 
Such a prospect is consistent with Ref.~\cite{Ho:2017bia} yet based on a distinguished nature of condensate.
In addition to the crystalline configuration, a fluid of baryonic vortex is also considerable, depending on the applicable crust of neutron stars. 
Decisive conclusions would be drawn after coupling the baryonic vortex with gravity and solving the equation of state.

On the other hand, our study caters to the interest in QCD phase diagram. 
Along the axis of $\mu_B$, at higher density regime where
%(compared to quark chemical potentials $ \mu_{u,d,s} \sim 400 \text{ MeV}$), 
strange quarks become relevant, we shall attempt to compare our spontaneous baryonic vortex with quark matter phases such as the crystalline color superconductivity~\cite{Alford:2000ze} and color-flavor-locking, in order to delve into the quark-hadron continuity.
In addition, within the hadronic regime, for the outlook of a more comprehensive digram of topological phases at finite $\mu_B$ and magnetic field, we shall further incorporate CSL, Skyrme crystal, domain-wall Skyrmions, mesonic quasicrystal etc, after figuring out the many-body system comprising of multiple baryonic vortices.

\begin{acknowledgments}
This work is supported in part by 
 JSPS KAKENHI [Grants No. JP22H01221] and the WPI program ``Sustainability with Knotted Chiral Meta Matter (SKCM$^2$)'' at Hiroshima University (MN).
\end{acknowledgments}
 
\section*{Appendix: Details of the Equation of Motion}

After some algebra, the explicit form of energy functional~\eqref{eq:H} before spatial integration with our Ansatz~\eqref{eq:ansatz} applied reads
\begin{align}
\mathcal{H}= & \frac{f_{\pi}^{2}}{2}\left\{ \left(\partial_{\rho}\alpha\right)^{2}+\left[\left(\frac{1}{\rho}+A_{\phi}\right)^{2}-\mu_{I}^{2}\right]\sin^{2}\alpha+\cos^{2}\alpha\left(\partial_{z}\varphi\right)^{2}\right\} \nonumber \\
 & +\frac{1}{2s^{2}}\left\{ \left(\partial_{\rho}\alpha\right)^{2}\cos^{2}\alpha\left(\partial_{z}\varphi\right)^{2}+\left[\left(\frac{1}{\rho}+A_{\phi}\right)^{2}-\mu_{I}^{2}\right]\sin^{2}\alpha\left[\left(\partial_{\rho}\alpha\right)^{2}+\cos^{2}\alpha\left(\partial_{z}\varphi\right)^{2}\right]\right\} \nonumber \\
 & -\frac{q}{4\pi^{2}} A_{\phi}\frac{\partial}{\partial\rho}\left(\mu_{I} \cos^{2}\alpha\right) \partial_{z}\varphi-\mu_{B}j_{B}^{0} +\frac{1}{2}\left[\frac{1}{\rho}\frac{\partial}{\partial\rho}\left(\rho A_{\phi}\right)\right]^{2}. 
\end{align}
Then with the direct corollary $\varphi=kz$, by the argument around Eq.~\eqref{eq:varphi}, we spell out the variational
principle of $H=\int d^3 x \mathcal{H}$ with respect to the transverse radius $\rho$, first for $\alpha$:
\begin{align}
 & \frac{\partial}{\partial\rho}\left\{ \rho\left\{ f_{\pi}^{2}+\frac{1}{s^{2}}\left(k^{2}\cos^{2}\alpha+\left[\left(\frac{1}{\rho}+A_{\phi}\right)^{2}-\mu_{I}^{2}\right]\sin^{2}\alpha\right)\right\} \partial_{\rho}\alpha\right\} \nonumber \\
= & \frac{\rho}{2}\Bigg\{ \left[\left(\frac{1}{\rho}+A_{\phi}\right)^{2}-\mu_{I}^{2}-k^{2}\right]\sin2\alpha\left[f_{\pi}^{2}+\frac{1}{s^{2}}\left(\partial_{\rho}\alpha\right)^{2}\right] \nonumber \\
& +\frac{1}{s^{2}}\sin2\alpha\cos2\alpha\left[\left(\frac{1}{\rho}+A_{\phi}\right)^{2}-\mu_{I}^{2}\right]k^{2} 
\Bigg\} 
\end{align}
Beware that the variation shall not be performed on the conserved
current $j_{B}^{\mu}$. So the WZW term gets involved in EOMs only
when we consider $\delta A_{\phi}$, but not $\delta\alpha$. Next, The
variational principle in terms of $A_{\phi}$ would reproduce the
Maxwell equation, $\partial_{\mu}F^{\mu\nu}=j_{Q}^{\nu}$:
\begin{align}
\frac{\partial}{\partial\rho}\left[\frac{1}{\rho}\frac{\partial}{\partial\rho}\left(\rho A_{\phi}\right)\right]
%& =-j_{Q}^{\phi} \nonumber \\
= & -\frac{qk}{4\pi^{2}}\frac{\partial}{\partial\rho}\left(\mu_{I}\cos^{2}\alpha\right) \nonumber\\ 
& +\left(\frac{1}{\rho}+A_{\phi}\right)\sin^{2}\alpha\left\{ f_{\pi}^{2}+\frac{1}{s^{2}}\left[\left(\partial_{\rho}\alpha\right)^{2}+k^{2}\cos^{2}\alpha\right]\right\},
\label{eq:eoma}
\end{align}
among which the first term on the r.h.s does originate from the WZW term and here
is the only portal where the WZW term enters the transverse EOMs. In other words, $\mu_B$ would not influence the profile under arbitrary $k$ but it takes part in determining the specific $k$ that leads to a minimized total free energy. 

\bibliographystyle{apsrev4-1}
\bibliography{baryonic_vortex}

\end{document}